\documentclass[12pt,a4paper]{elsart}
\usepackage{ifthen,graphics}
\usepackage{color}
\usepackage{cite}
\usepackage[colorlinks]{hyperref}
\usepackage{epsfig}

\catcode`@=11 %%At signs are letters
\newdimen\z@ \z@=0pt % can be used both for 0pt and 0
\newskip\z@skip \z@skip=0pt plus0pt minus0pt
\def\m@th{\mathsurround=\z@}
\def\ialign{\everycr{}\tabskip\z@skip\halign} % initialized \halign
\def\eqalign#1{\null\,\vcenter{\openup\jot\m@th
  \ialign{\strut\hfil$\displaystyle{##}$&$\displaystyle{{}##}$\hfil
      \crcr#1\crcr}}\,}
\catcode`@=12 % at signs are no longer letters

\def\bye{\bibliographystyle{elsart-num}
\bibliography{note}

\end{document}
}
\def \gev  {{\rm \,Ge\kern-0.125em V}}
\def \mev  {{\rm \,Me\kern-0.125em V}}
\def \kev  {{\rm \,ke\kern-0.125em V}}
\def \ev   {{\rm \,e\kern-0.125em V}}

\newcommand{\pvec}{\ensuremath{\mathbf{p}}}
\newcommand{\xvec}{\ensuremath{\mathbf{x}}}
\newcommand{\rxy}{\ensuremath{r_{xy}}}

\newcommand{\dTCA}{\ensuremath{\it{d}_{\mathrm{TC}}}}
\newcommand{\dtTCA}{\ensuremath{\it{d}_{\mathrm{TC}\bot}}}

\newcommand{\xc}{\ensuremath{\mathbf{x}_\mathrm{c}}}
\newcommand{\pc}{\ensuremath{\mathbf{p}_\mathrm{c}}}
\newcommand{\lc}{\ensuremath{l_\mathrm{c}}}
\newcommand{\dc}{\ensuremath{d_\mathrm{c}}}
\newcommand{\fphat}{\mbox{$\hat{f}_+(t)$}}
\newcommand{\fpha}{\mbox{$\hat{f}_+$}}

\def\ks{\ifm{K_S}} 
\def\kl{\ifm{K_L}}

\newcommand{\lam}{\mbox{$\lambda_+$}}
\newcommand{\lamp}{\mbox{$\lambda'_+$}}
\newcommand{\lampp}{\mbox{$\lambda''_+$}}
\newcommand{\klpen}{\mbox{$K_{L}\to\pi^{\pm}e^{\mp}\nu$}}
\newcommand{\klele}{\mbox{$K_{L}\to\pi^+e^-\overline{\nu}$}}
\newcommand{\klpos}{\mbox{$K_{L}\to\pi^-e^+\nu$}}
\newcommand{\klpmn}{\mbox{$K_{L}\to\pi\mu\nu$}}
\newcommand{\klpln}{\mbox{$K_{L}\to\pi^{\pm}\ell^{\mp}\nu$}}
\newcommand{\klpp}{\mbox{$K_{L}\to\pi^+\pi^- $}}
\newcommand{\kspp}{\mbox{$K_{S}\to\pi^+\pi^- $}}
\newcommand{\klppp}{\mbox{$K_{L}\to\pi^+\pi^-\pi^0$}}

\newcommand{\miss}{\mbox{$E_{miss}-p_{miss}$}}
\newcommand{\semil}{\mbox{$K_{e3}$}}

\newcommand{\tof}{\mbox{TOF}}
\newcommand{\kloe}{\mbox{KLOE}}

\newcommand{\geanfi}{\mbox{GEANFI}}
\newcommand{\ith}{\ensuremath{i^\mathrm{th}}}
\newcommand{\jth}{\ensuremath{j^\mathrm{th}}}

\newcommand{\ie}{{\em i.e.}}

\newcommand{\Eq}[1]{Eq.~(\ref{#1})}  
\newcommand{\Fig}[1]{Fig.~\ref{#1}}
\newcommand{\Ref}[1]{Ref.~\citen{#1}}

\newcommand{\Tab}[1]{Table~\ref{#1}}
\newcommand{\aff}[2]{Dipartimento di Fisica dell'Universit\`a #1 e Sezione INFN, #2, Italy.}
\newcommand{\affd}[1]{Dipartimento di Fisica dell'Universit\`a e Sezione INFN, #1, Italy.}
\let\cl=\centerline
\def\figbox#1;#2;{\parbox{#2cm}{%
\vglue3mm\epsfig{file=#1.eps,width=#2cm}\vglue3mm}}
\def\figboxc#1;#2;{\cl{\figbox #1;#2;}}
\def\ifm#1{\relax\ifmmode#1\else$#1$\fi}
\def\lam{\ifm{\lambda_+}}
  
\font\euler=eufm10 at 12pt
\def\Ma{\hbox{\euler M}}
\begin{document}
\begin{frontmatter}
  \begin{center} 
    {\Large \bf Measurement of the Form-Factor Slopes for the Decay \klpen\ with the \kloe\ Detector}
  \end{center}
  \pagestyle{plain}
\collab{The KLOE Collaboration}
\author[Na]{F.~Ambrosino},
\author[Frascati]{A.~Antonelli}
\author[Frascati]{M.~Antonelli\corauthref{cor1}},
\author[Roma3]{C.~Bacci},
\author[Karlsruhe]{P.~Beltrame},
\author[Frascati]{G.~Bencivenni},
\author[Frascati]{S.~Bertolucci},
\author[Roma1]{C.~Bini},
\author[Frascati]{C.~Bloise},
\author[Roma1]{V.~Bocci},
\author[Frascati]{F.~Bossi},
\author[Virginia]{D.~Bowring},
\author[Roma3]{P.~Branchini},
\author[Roma1]{R.~Caloi},
\author[Frascati]{P.~Campana},
\author[Frascati]{G.~Capon},
\author[Na]{T.~Capussela},
\author[Roma3]{F.~Ceradini},
\author[Frascati]{S.~Chi},
\author[Na]{G.~Chiefari},
\author[Frascati]{P.~Ciambrone},
\author[Virginia]{S.~Conetti},
\author[Frascati]{E.~De~Lucia},
\author[Roma1]{A.~De~Santis},
\author[Frascati]{P.~De~Simone},
\author[Roma1]{G.~De~Zorzi},
\author[Frascati]{S.~Dell'Agnello},
\author[Karlsruhe]{A.~Denig},
\author[Roma1]{A.~Di~Domenico},
\author[Na]{C.~Di~Donato},
\author[Pisa]{S.~Di~Falco},
\author[Roma3]{B.~Di~Micco},
\author[Na]{A.~Doria},
\author[Frascati]{M.~Dreucci\corauthref{cor2}},
\author[Frascati]{G.~Felici},
\author[Karlsruhe]{A.~Ferrari},
\author[Frascati]{M.~L.~Ferrer},
\author[Frascati]{G.~Finocchiaro},
\author[Roma1]{S.~Fiore},
\author[Frascati]{C.~Forti},
\author[Roma1]{P.~Franzini},
\author[Frascati]{C.~Gatti\corauthref{cor3}},
\author[Roma1]{P.~Gauzzi},
\author[Frascati]{S.~Giovannella},
\author[Lecce]{E.~Gorini},
\author[Roma3]{E.~Graziani},
\author[Pisa]{M.~Incagli},
\author[Karlsruhe]{W.~Kluge},
\author[Moscow]{V.~Kulikov},
\author[Roma1]{F.~Lacava},
\author[Frascati]{G.~Lanfranchi},
\author[Frascati,StonyBrook]{J.~Lee-Franzini},
\author[Karlsruhe]{D.~Leone},
\author[Frascati]{M.~Martini},
\author[Na]{P.~Massarotti},
\author[Frascati]{W.~Mei},
\author[Na]{S.~Meola},
\author[Frascati]{S.~Miscetti},
\author[Frascati]{M.~Moulson},
\author[Karlsruhe]{S.~M\"uller},
\author[Frascati]{F.~Murtas},
\author[Na]{M.~Napolitano},
\author[Roma3]{F.~Nguyen},
\author[Frascati]{M.~Palutan},
\author[Roma1]{E.~Pasqualucci},
\author[Roma3]{A.~Passeri},
\author[Frascati,Energ]{V.~Patera},
\author[Na]{F.~Perfetto},
\author[Roma1]{L.~Pontecorvo},
\author[Lecce]{M.~Primavera},
\author[Frascati]{P.~Santangelo},
\author[Roma2]{E.~Santovetti},
\author[Na]{G.~Saracino},
\author[Frascati]{B.~Sciascia},
\author[Frascati,Energ]{A.~Sciubba},
\author[Pisa]{F.~Scuri},
\author[Frascati]{I.~Sfiligoi},
\author[Frascati]{T.~Spadaro},
\author[Roma1]{M.~Testa},
\author[Roma3]{L.~Tortora},
\author[Frascati]{P.~Valente},
\author[Karlsruhe]{B.~Valeriani},
\author[Frascati]{G.~Venanzoni},
\author[Roma1]{S.~Veneziano},
\author[Lecce]{A.~Ventura},
\author[Karlsruhe]{R.~Versaci},
\author[Frascati,Beijing]{G.~Xu},
%%%
\address[Virginia]{Physics Department, University of Virginia, Charlottesville, VA, USA.}
\address[Frascati]{Laboratori Nazionali di Frascati dell'INFN, Frascati, Italy.}
\address[Karlsruhe]{Institut f\"ur Experimentelle Kernphysik, Universit\"at Karlsruhe, Germany.}
\address[Lecce]{\affd{Lecce}}
\address[Na]{Dipartimento di Scienze Fisiche dell'Universit\`a ``Federico II'' e Sezione INFN, Napoli, Italy}
\address[Energ]{Dipartimento di Energetica dell'Universit\`a ``La Sapienza'', Roma, Italy.}
\address[Roma1]{\aff{``La Sapienza''}{Roma}}
\address[Roma2]{\aff{``Tor Vergata''}{Roma}}
\address[Roma3]{\aff{``Roma Tre''}{Roma}}
\address[Pisa]{\affd{Pisa}}
\address[StonyBrook]{Physics Department, State University of New York at Stony Brook, NY, USA.}
\address[Beijing]{Permanent address: Institute of High Energy Physics, CAS, Beijing, China.}
\address[Moscow]{Permanent address: Institute for Theoretical and Experimental Physics, Moscow, Russia.}
\begin{flushleft}
\corauth[cor1]{cor1}{\small $^1$ Corresponding author: Mario Antonelli
INFN - LNF, Casella postale 13, 00044 Frascati (Roma), 
Italy; tel. +39-06-94032728, e-mail mario.antonelli@lnf.infn.it}
\end{flushleft}
\begin{flushleft}
\corauth[cor2]{cor2}{\small $^2$ Corresponding author: Marco Dreucci
INFN - LNF, Casella postale 13, 00044 Frascati (Roma), 
Italy; tel. +39-06-94032696, e-mail marco.dreucci@lnf.infn.it}
\end{flushleft}
\begin{flushleft}
\corauth[cor3]{cor3}{\small $^3$ Corresponding author: Claudio Gatti
INFN - LNF, Casella postale 13, 00044 Frascati (Roma), 
Italy; tel. +39-06-94032727, e-mail claudio.gatti@lnf.infn.it}
\end{flushleft}

%

%+++++++++++++++
\begin{abstract}
%++++++++++++++++
 We present a measurement of the $K$-$\pi$ vector current form-factor parameters for the decay 
 \klpen. We use 328 pb$^{-1}$ of data collected in 2001 and 2002, corresponding to
 $\sim $ 2 million \semil\ events.  
 Measurements of semileptonic form factors provide
 information about the dynamics of the strong interaction and are necessary for evaluation 
 of the phase-space integral $I^e_K$ needed to measure the CKM matrix element $|V_{us}|$ 
 for \klpen\ decays. 
 Our result is $\lam=(28.6\pm0.5\pm0.4)10^{-3}$ for a linear fit,
 and $\lamp=(25.5\pm1.5\pm1.0)10^{-3}$, $\lampp=(1.4\pm0.7\pm0.4)10^{-3}$ for a quadratic
 fit.
\vspace{.5cm}\\
{\it key words:}~ke3 form factor 
\end{abstract} 
\end{frontmatter}
\newpage
\setcounter{page}{1}

%+++++++++++++++++++++++
\section{Introduction}
%++++++++++++++++++++++++
%
Semileptonic kaon decays, \klpln, (\Fig{fig:diag}) offer possibly the cleanest way to obtain  an 
accurate value of the Cabibbo angle, or better, $V_{us}$. Since $K\to \pi$ is a $0^-\to 0^-$ transition, 
only the vector part of the weak current has a nonvanishing contribution.  The transition is therefore 
protected by the Ademollo-Gatto theorem against SU(3) breaking corrections to lowest order. At present, 
the largest uncertainty in calculating $V_{us}$ from the decay rate, is due to the difficulties in 
computing the matrix element $\langle\pi|j_{\mu}|K\rangle$.
%---------------------------------------------------------------------
\begin{figure}[ht]
\figboxc ke3-n;4;  
\caption{Amplitude for \klpln. The gray region indicates the $K\to\pi W$ vertex structure. }
  \label{fig:diag}
\end{figure}
%---------------------------------------------------------------------
In the electron mode \klpen\ only one form factor is involved. In the following we will use the notation 
shown in \Fig{fig:diag}, in which $P$, $p$, $k$ and $k'$ are the kaon, pion, electron and neutrino momenta,
respectively; $m$ is the mass of the charged pion and $M$ that of the neutral kaon. Terms in $(P-p)_{\mu}$ 
that acquire factor of $m_e$ are neglected. Therefore:\vglue-5mm
%-----
\begin{eqnarray*}
  \label{eq:hc1} 
  \langle \pi|J^V_{\mu}|K \rangle &=& f_+(t) (P+p)_{\mu} 
\end{eqnarray*}
%-----------------------------------------------------------------------
We replace the form factor above with $f_+(0)\,\fphat$, where $t=(P-p)^2=(k+k')^2=M^2+m^2-2ME_\pi$ is the
 only {\it{L}}-invariant variable and $\fpha(0)=1$. 
The form factor is dominated by the vector $K$-$\pi$ resonances, the closest being the $K^*(892)$. Note 
that for $t>0$, \fphat\ $>1$. The presence of the form factor increases the value of the phase-space 
integral and the decay rate.
The natural form for $\hat f(t)$ is
\begin{equation}
\fphat ={M_V^2\over M_V^2-t}.
  \label{eq:pole}
\end{equation}
It is also customary to expand the form factor as
%------------------------------------------------------------------------------------------
\begin{equation}
  \fphat = 1 + \lamp~\frac{t}{m^2}+\frac{1}{2}\;\lampp\,\left(\frac{t}{m^2}\right)^2\,+\dots
  \label{eq:ff2}
\end{equation}
%------------------------------------------------------------------------------------------
In the following we retain linear and quadratic terms. Note that the expansion of the pole form above gives
 $\lamp=(m/M_V)^2$ and $\lampp=2\,\lamp^2$. From $P+p=2P-(k+k')$, and neglecting the $k+k'$ term which is 
also proportional to $m_e$, the amplitude is:
%-----------------------------------------------------------------------------------
\begin{equation}
  \label{eq:hc2} 
    \Ma = \langle\pi e\nu|H_W|K\rangle = 2\frac{V_{us} G_F f_+(0)}{\sqrt 2}
    P_{\mu} \bar u(k) \gamma^{\mu}(1-\gamma^5)u(k') \fphat. 
\end{equation}
%-----
\def\lo{\lower1mm}
Squaring, summing over spins, and integrating over all variables but the pion energy, we obtain the pion 
spectrum
$$g(z)\propto\,\left( z^2 - 4\,\alpha  \right)^{3\over2}\, \left(1 +{\lambda_+'\over\alpha}\, \left(\xi-z\right)+       {\lambda_+''\over2\,\alpha^2}\,\left(\xi-z\right)^2\right)^2$$
where $\alpha=m^2/M^2,\ \xi=1+\alpha$ and $z=2\,E_\pi/M$ is the normalized pion energy. The spectrum can 
also be written in terms of $t=m^2(\xi-z)/\alpha$ as
%---------------
\begin{equation}
  \eqalign{h(t)&\,\propto\,\lambda^{3/2} \fphat^2\cr
    \lambda&=t^{\,2}-2\,t\,(M^{\,2} + 2{\,}m^{\,2}) + (M^{\,2}-m^{\,2})^{\,2} }
    \label{eq:rate}
\end{equation}

%=========================================%
\section{The KLOE detector}
\label{sec:kloedet}
%=========================================%
The KLOE detector consists of a large, cylindrical drift chamber (DC), surrounded by a 
lead/scintillating-fiber electromagnetic calorimeter (EMC). A superconducting coil around the calorimeter 
provides a 0.52 T field. The drift chamber \cite{KLOE:DC} is 4~m in diameter and 3.3~m long.
The momentum resolution is $\sigma_{p_{\perp}}/p_{\perp}\approx 0.4\%$. 
Two-track vertices are reconstructed with a spatial resolution of $\sim$ 3~mm. 
The calorimeter \cite{KLOE:EmC} is divided into a barrel and two endcaps. It covers 98\% of the solid 
angle. Cells close in time and space are grouped into calorimeter clusters. The energy and time resolutions 
are $\sigma_E/E = 5.7\%/\sqrt{E\ {\rm(GeV)}}$ and $\sigma_t = 57\ {\rm ps}/\sqrt{E\ {\rm(GeV)}}\oplus100\ {\rm ps}$, 
respectively.
The KLOE trigger \cite{KLOE:trig} uses calorimeter and chamber information. For this 
analysis, only the calorimeter signals are used. Two energy deposition above threshold ($E>50$ MeV for the 
barrel and  $E>150$ MeV for the endcaps) are required. Recognition and rejection of cosmic-ray events is 
also performed at the trigger level. Events with two energy deposition above a 30 MeV threshold in the 
outermost calorimeter plane are rejected.

%*******************
\section{Analysis}
%*******************
The 328 pb$^{-1}$ of 2001-2002 data used in this analysis~\cite{KLOE:ffnote}, 
is divided into 14 periods of about 25 pb$^{-1}$/period.
For each data period we have a corresponding  sample of Monte Carlo events 
with approximately the same statistics.

Candidate \kl\ events are tagged by the presence of a \kspp\ decay. 
The \kl\ tagging algorithm is fully described in~\cite{KLOE:brlnote} and~\cite{KLOE:brl}.
The \kl\ momentum, $p_{\kl}$, is obtained from the kinematics of the 
$\phi\to\ks\kl$ decay, using the \ks\ direction reconstructed from the measured 
momenta of the decay tracks and the known value of $\pvec_{\phi}$. 
The resolution  is dominated by the beam-energy spread, and amounts to about 
0.8 MeV/c. The position of the $\phi$\ production point, $\xvec_\phi$, is determined 
as the point of closest approach of the \ks, propagated backward
from the \ks\ vertex, to the beam line. The \kl\ line of flight 
({\em tagging line}) is then constructed from the \kl\ momentum,  
$\pvec_{\kl} = \pvec_\phi-\pvec_{\ks}$, and the position of the production vertex, 
$\xvec_\phi$.
 
The efficiency of the tagging procedure depends slightly on the evolution of the \kl,
mainly because the trigger  efficiency depends on the \kl\ behavior.
To identify events in which the \ks\ by itself satisfies the calorimeter 
trigger, we require the presence of two clusters from 
the $\kspp$ decay associated with fired trigger sectors ({\em{autotrigger}}).
The value of the tagging efficiency obtained from
Monte Carlo is about 40 \% and  is independent of $t$ to within 0.4\%.

All tracks in the chamber, after removal of those from the \ks\ 
decay and their descendants, are extrapolated to their points of closest
approach to the tagging line. 
For each track candidate, we evaluate the point of closest approach to the 
tagging line, \xc, and the distance of closest approach, \dc. The
momentum \pc\ of the track at \xc\ and the extrapolation length \lc\ are also
computed. Tracks satisfying $\dc< a \rxy + b$, with $a=0.03$ and $b=3$ cm, and 
$-20$ cm $<\lc<25$ cm are accepted as \kl\ decay products, where \rxy\ is the
distance of the vertex from the origin in the transverse plane. For each sign of 
charge we consider the track with the smallest value of \dc to be associated to the \kl\ 
decay. Starting from these track candidates a vertex is reconstructed.
The combined tracking and vertexing efficiency for \semil\ decays is about 54\%.
It is determined from data as described in \Ref{KLOE:brlnote}. An event is retained if 
the vertex is in the fiducial volume $35<\rxy<150$ cm and $|z|<120$ cm.

To remove background from \klppp\ and \klpp\ decays with minimal efficiency loss, we 
apply loose kinematic cuts: assuming the two tracks to have the pion mass,  
we require $E^2_{miss}-p^2_{miss}-M^2_{\pi^0}<-5000~\mev^2$ and 
$\sqrt{E^2_{miss}+p^2_{miss}} > 10\mev$, where $E_{miss}$ and $p_{miss}$
are the missing energy and momentum, respectively.
A large amount of background from \klpmn\ decays is rejected using
$\Delta_{\pi\mu}$, the lesser value of $|\miss|$ calculated in the two hypotheses, 
$\pi\mu$ or $\mu\pi$.
We retain events only if this variable is greater than 10\mev.
After the kinematic cuts described above, the efficiency for the signal is about 96\%.

These kinematic criteria do not provide enaugh suppression of the background from 
\klpen\ decays with incorrect charge assignment and from \klpmn\ decays.
We make use of time-of-flight (\tof) information to further reduce the
contamination.   

For the purpose of track-to-cluster association, 
we define two quantities related to the distance between the track, extrapolated 
to the entry point of the calorimeter, and the closest 
cluster: \dTCA, the distance from the extrapolated entry point on the calorimeter 
to the cluster centroid and \dtTCA, the component of this distance in the 
plane transverse to the momentum of the track at the entry position.
We only consider clusters with \dtTCA\ $<$ 30~cm and $E>$50~\mev.

\begin{figure}[ht]
  \centering
  \psfig{figure=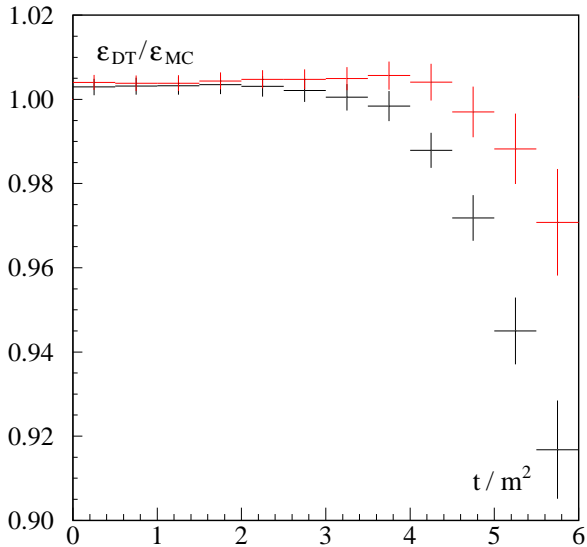,width=8.0cm}
  \caption{Cluster efficiency correction for pions ($\pi^-$ in red, $\pi^+$ in
    black) on the end cap from control sample. The plot refers to a single period 
    of data taking}
  \label{clupit}
\end{figure}

We evaluate the cluster efficiency using the Monte Carlo, and correct it
with the ratio of data and Monte Carlo efficiencies obtained from control samples.
A sample of \semil\ events with a purity of 99.5\% is selected by means of kinematics and 
independent calorimeter information.
Figure~\ref{clupit} shows the corrections as a function of $t$ obtained for a
single run period.
It is worth emphasizing that if this correction were not taken into account, the
effect on \lamp\ would be large (about 20\%) and would produce different results for  
each charge (about 15\%). For this reason, the analysis is performed separately for 
each charge. The comparison of the two results provides a first check of the validity of 
the corrections.

For each \kl\ decay track with an associated cluster, we define the variable: 
$\Delta t_i = t_{\rm cl}-t_{i} ,~(i = \pi,~ e) $ in which $t_{\rm cl}$ is the cluster time 
and $t_{i}$ is the expected time of flight, evaluated according to a well-defined mass 
hypothesis. The evaluating of $t_{i}$ includes the propagation time from 
the entry point to the cluster centroid~\cite{KLOE:kssemi}.
We determine the $e^+e^-$ collision time, $t_0$, using the clusters from the \ks.

\begin{figure}[ht]
  \centering
  \psfig{figure=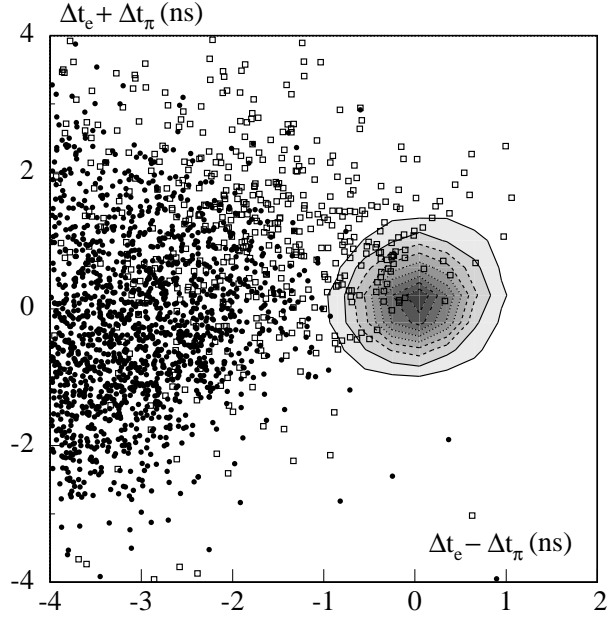,width=8.0cm}
  \caption{Monte Carlo distribution of $\Delta t_e+\Delta t_{\pi}$ versus 
    $\Delta t_e-\Delta t_{\pi}$.
    Signal (gray scale), background from opposite sign \klpen\ (black dot), background from 
    same sign \klpmn\ (empty box).}
  \label{cuttof}
\end{figure}

An effective way to select the correct mass assignment, $\pi e$ or $e\pi$,
is obtained by choosing the lesser of $|\Delta t_{\pi^+}-\Delta t_{e^-}|$ and
$|\Delta t_{\pi^-}-\Delta t_{e^+}|$.
After the mass assignment has been made, we consider the variables
$\Delta t_{\pi}+\Delta t_e$ and $\Delta t_{\pi}-\Delta t_e$. 
These are shown in~\Fig{cuttof} for signal and background Monte Carlo events.
We select the signal by using a $2\sigma$ cut, where the resolution 
$\sigma\simeq 0.5~\mbox{ns}$.
After the \tof\ cut we have a contamination of $\sim$0.7\%, almost entirely due to 
\klpmn\ decays.

\begin{figure}[ht]
  \centering
  \begin{tabular}{cc}
     \psfig{figure=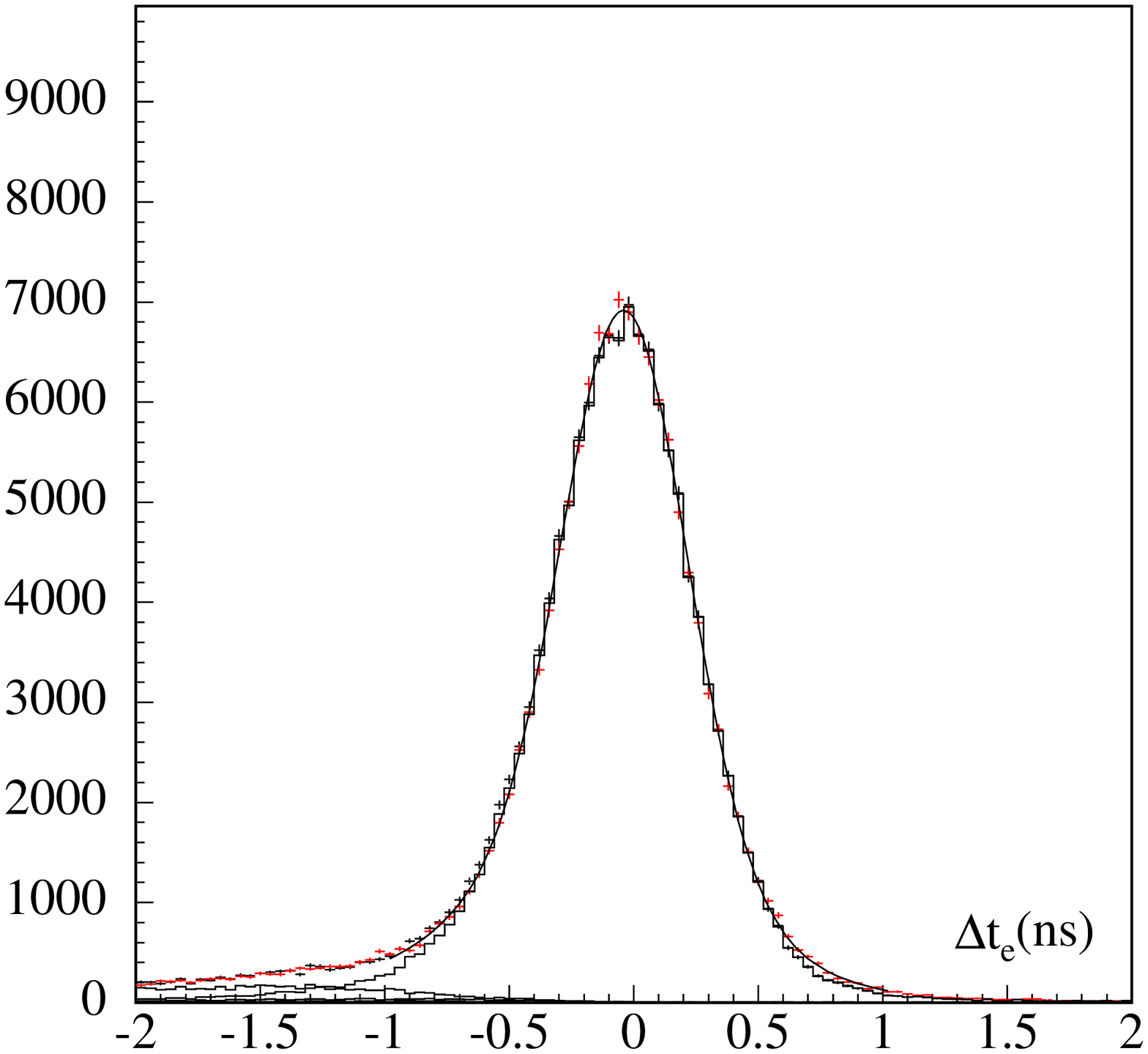,width=6.2cm}
     &
     \psfig{figure=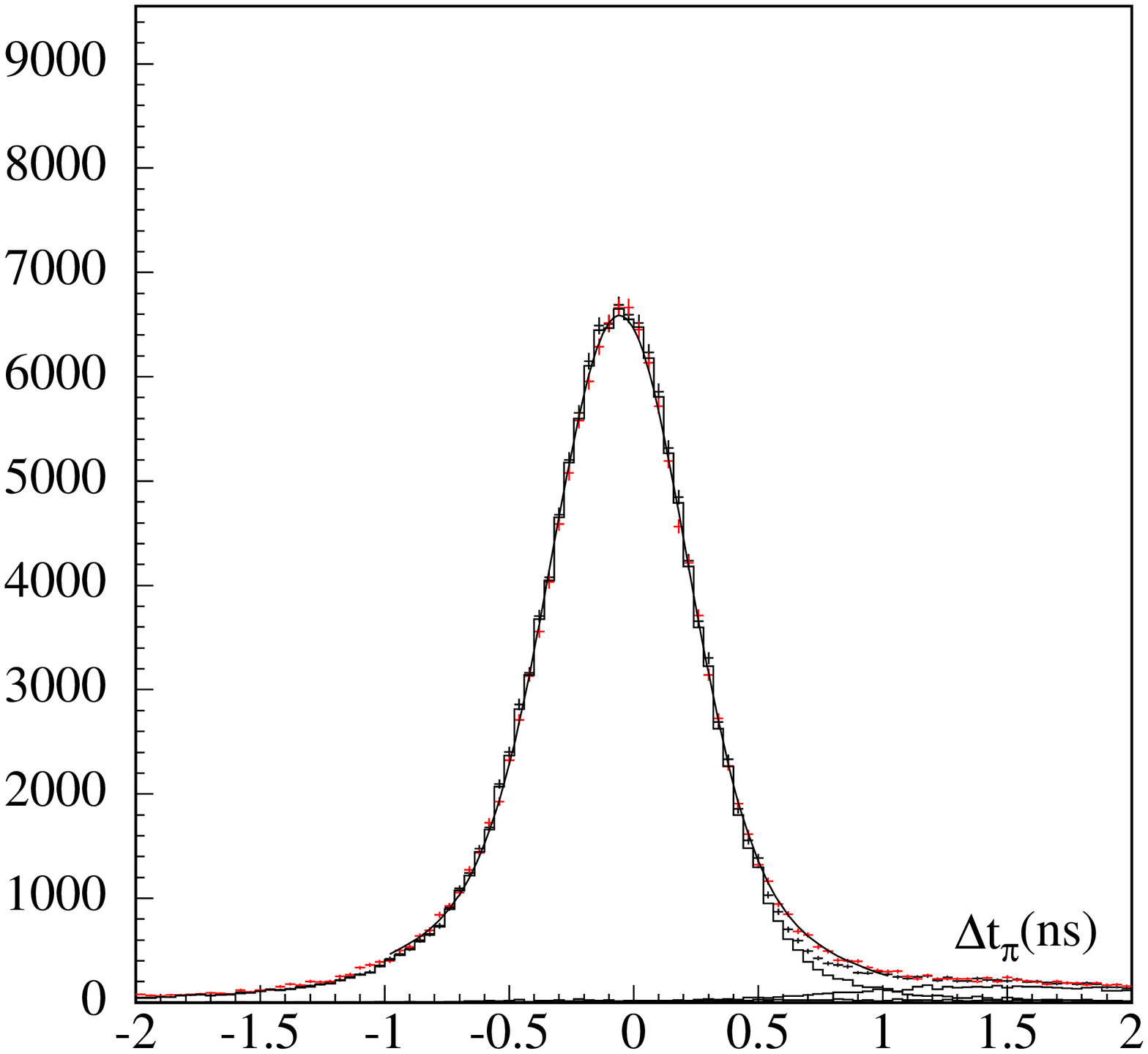,width=6.2cm}\\
     (a) & (b)\\
  \end{tabular}
  \caption{$\Delta t_i$ for electron (a) and pion (b) for data (black) and Monte Carlo (red)}
  \label{tof}
\end{figure}

We take the \tof\ efficiency from the Monte Carlo after correcting the time 
response of the calorimeter using data control samples~\cite{KLOE:kssemi}.
The quality of this correction can be checked by comparing
the data and Monte Carlo distributions for $\Delta t_e$ and $\Delta t_{\pi}$  
shown in \Fig{tof}, obtained from the same control sample used for the cluster efficiency.

We measure the form-factor slope parameters by fitting the distribution of the selected 
events in $t/m^2$. We modify the kinematic range of $t/m^2$, varying from 
$(m_e/m)^2 \sim 10^{-5}$ up to $(M-m)^2/m^2 \sim$ 6.8, to [-0.5,6],
to take into account the smearing effect at $t\sim 0$ and the low statistics at high values 
of $t$. 
After subtracting the residual background as estimated from Monte Carlo, we perform the fit
using the following formula:
%---------------------------------------------------------------------------------------------
\begin{equation}
  \label{fitfunc}
  \frac{dN}{dt}(i)=N_0\sum_{j=1}^{20} A(i,j)\times~\rho(j,\lamp,\lampp)\times~\epsilon_{tot}(j)
  \times~F_{FSR}(j)
\end{equation}
%---------------------------------------------------------------------------------------------
where $\rho(j,\lamp,\lampp)$ is the three-body differential decay width as defined in
\Eq{eq:rate}, and $A(i,j)$ is the probability that an event with true value of $t/m^2$ 
in the \jth\ bin has a reconstructed value in the \ith\ bin.
The chosen bin size is 0.5, which corresponds to about 1.6 $\sigma_t$, where $\sigma_t$
is the resolution in $t/m^2$.

The total efficiency, $\epsilon_{\rm{tot}}(t)$, takes into account the acceptance and the 
efficiency of the analysis cuts. 
$F_{\rm{FSR}}$ is the correction for final-state radiation.
It is evaluated using the \kloe\  Monte Carlo  simulation, \geanfi\ ~\cite{KLOE:offline},
where FSR processes are simulated according the procedures described in \Ref{KLOE:gattip}.
FSR affects the $t$-distribution mainly for high-energy pions,
\ie\ for low $t$, where the correction is 3-5\%.   
The slopes \lamp\ and \lampp\ are free parameters in the fit while the $N_0$
constant is the total number of signal events.

%+++++++++++++++++++++++++++++++++
\section{Systematic uncertainties}
%+++++++++++++++++++++++++++++++++
The systematic errors due to the evaluation of corrections, data-Monte Carlo inconsistencies, 
result stability, momentum miscalibration, and background contamination are summarized 
in \Tab{tabsyst}.

\begin{table}[hbt]
  \begin{center}
    \begin{tabular}{|c|c|c|c|}
      \hline 
    &    Linear fit &   \multicolumn{2}{|c|}{ Quadratic fit}     \\
      \hline
   Source & $\delta\lam\times 10^{3}$ & $\delta\lamp\times 10^{3}$ & $\delta\lampp\times 10^{3}$\\
      \hline
      Tagging                 &    0.14   &  0.18  & 0.02   \\
      Tracking and vertexing  &    0.16   &  0.22  & 0.18   \\  
      Clustering              &    0.07   &  0.24  & 0.13   \\
      Time-of-flight          &    0.29   &  0.87  & 0.27   \\
      Background              &    0.08   &  0.16  & 0.03   \\
      Momentum-scale          &    0.06   &  0.05  & 0.05   \\
      Momentum-resolution     &    0.17   &  0.22  & 0.19   \\
      \hline
      Total systematic        &    0.42   &  0.98  & 0.39   \\
      \hline
    \end{tabular}  
    \caption{Summary of systematic uncertainties on \lam, \lamp and \lampp.}
    \label{tabsyst}    
  \end{center}
\end{table}
%

% TAGGING
We evaluate the systematic uncertainty of the tagging efficiency  
by repeating the measurement using a tagging algorithm without the requirement of 
the autotrigger. We observe a change of 0.14$\times10^{-3}$ for \lam\ in the case
of the linear fit, and changes of 0.18$\times10^{-3}$ and 0.02$\times10^{-3}$
for \lamp\ and \lampp\ respectively for the quadratic fit.

% TRACKING
We evaluate the systematic uncertainties on the tracking efficiency corrections by 
checking stability of the result when the track selection criteria are modified.
We establish the validity of the method by comparing the efficiencies from data 
and Monte Carlo control samples, and from the Monte Carlo truth~\cite{KLOE:kssemi}.
The uncertainty on the tracking efficiency correction is dominated by sample statistics 
and by the variation of the results observed using different criteria to identify tracks 
from \kl\ decays.
The correction is run-period dependent; its statistical error is taken into account in the 
fit. We study the effect of differences in the resolution with which the variable
\dc\ is reconstructed in data and in Monte Carlo events, and the possible bias introduced 
in the selection of the control sample, by varying the values of the 
cuts made on this variable when associating tracks to \kl\ vertices.
For each variation, corresponding to a maximal change of the tracking efficiency
of about $\pm$15\%, we evaluate the complete tracking-efficiency correction 
and measure the slope parameters.
We observe a change of 0.16$\times10^{-3}$ for \lam\ in the case 
of the linear fit, and changes of  0.22$\times10^{-3}$ and 0.18$\times10^{-3}$ 
for \lamp\ and \lampp\ , respectively, for the quadratic fit.
We find a smaller uncertainty by comparing the efficiencies 
from data and Monte Carlo control samples, and Monte Carlo truth. 
However, we conservatively assume the systematic uncertainty to be given by the changes 
in the result observed by varying the cut on \dc.

% CLUSTERING
We evaluate the systematic uncertainties on the clustering efficiency corrections 
by checking stability of the result when the track-to-cluster association criteria are 
modified. In this case as well, the uncertainties on the clustering efficiency corrections 
are dominated by sample statistics and by the variation of the results observed using 
different criteria for the track-to-cluster association. 
The correction is run-period dependent; we take into account its statistical error in the fit.
The most effective variable in the definition of track-to-cluster association
is the transverse distance, \dtTCA. We vary the cut on  \dtTCA\  in a wide range from 7 cm 
to 30 cm, corresponding to a change in efficiency of about 17\%.
For each configuration, we obtain the complete track extrapolation and clustering 
efficiency correction and use it to evaluate the slopes.
We observe a variation of 0.07$\times10^{-3}$ for \lam\ in the case
of the linear fit, and variations of 0.24$\times10^{-3}$ and 0.13$\times10^{-3}$
for \lamp\ and \lampp\ , respectively, in the case of the quadratic fit.
We find a comparable uncertainty for \lamp\ and for \lampp\ by comparing the efficiencies from 
data and Monte Carlo control samples, and the Monte Carlo truth.

% TOF
We study the uncertainty on the Monte Carlo efficiency of the \tof\ selection
procedure by measuring it using a pure \semil\ control sample,
and using the ratio of data and Monte Carlo efficiencies estimated in this way
as a correction.\footnote{The \tof\ corrections cannot be used directly in the analysis  
because of the correlation between the energy response in the
calorimeter and the \tof.}
The control sample is selected using tighter kinematic cuts and 
the calorimeter particle identification described in \Ref{KLOE:brlnote}. 
The  contamination of the control sample amounts to 0.4\%.
When applying the correction, we find a  change in the result of 0.29$\times10^{-3}$  
for \lam\ in the case of the linear fit, and changes of  
0.87$\times10^{-3}$ and 0.27$\times10^{-3}$
for \lamp\ and \lampp\ , respectively, in the case of the quadratic fit. 
These variations are well within the statistical uncertainties.

\begin{figure}[ht]
   \centering
   \psfig{figure=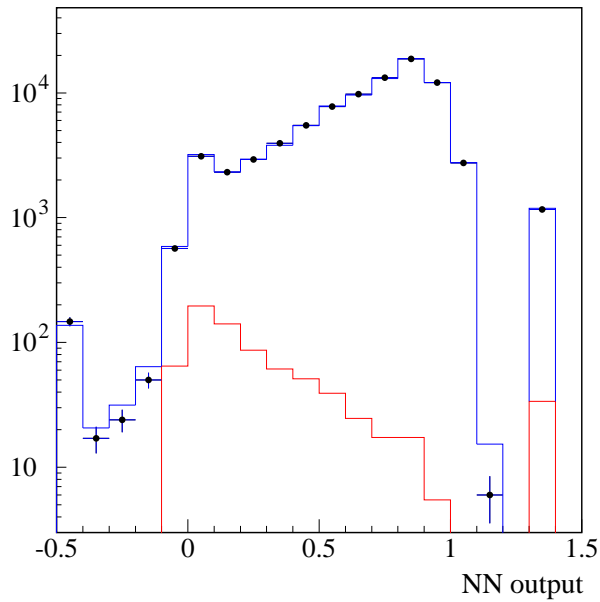,width=8.0cm}
   \caption{NN output distribution for data (dots), Monte Carlo signal
     (black solid line) and background (red solid line).}
   \label{fig:nnout}
 \end{figure}

% BACKGROUND
We study the uncertainties on the background evaluation by repeating the measurement on a 
sample with reduced background contamination.
This is achived by identifying the electron using the energy deposition in the calorimeter 
combined with a neural network (NN). In \Fig{fig:nnout} we show the distribution of the 
NN output for the sample used in the analysis. Requiring the value of the NN output to be 
greater than 0.4, we reduce the background contamination by about a factor of three.
The differences in the result obtained with this cut are 0.08$\times10^{-3}$ for \lam\ in 
the case of the linear fit, and 0.16$\times10^{-3}$ and 0.03$\times10^{-3}$ for \lamp\ and 
\lampp\ , respectively, in the case of the quadratic fit.

% P-SCALE and P-RESOLUTION
The effect of the momentum scale uncertainty and the momentum resolution have also been 
considered. We find the following relations by changing the momentum scale:
\[
\frac{\delta \lamp}{\lamp}\simeq -2 \frac{\delta p}{p},~~~~ \frac{\delta \lampp}{\lampp}\simeq -4\frac{\delta p}{p}
\]
We conservatively assume a momentum scale uncertainty of 0.1\%, which is much greater 
than the value obtained from a dedicated analysis \cite{memo:magnfield}. This
translates into a change of 0.06$\times10^{-3}$ for \lam\ in the case of the linear fit, and 
changes of 0.05$\times10^{-3}$ and 0.05$\times10^{-3}$ for \lamp\ and \lampp\ , respectively, 
in the case of the quadratic fit.

We investigate the effect of the momentum resolution by changing the value of the 
resolution on $t/m^2$. A variation of 3\% worsens the fit quality, giving a $\chi^2$ 
probability variation of one standard deviation. 
The corresponding absolute changes are 0.17$\times10^{-3}$ for 
\lam\ in the case of the linear fit, and 0.22$\times10^{-3}$ and 0.19$\times10^{-3}$ for 
\lamp\ and \lampp\ in the case of quadratic fit. 
Varying the resolution of $t/m^{2}$ by 
a larger amount (6\%) gives an unacceptable $\chi^2$ probability, about $10^{-9}$, while 
nearly the same variations for the fit parameters are observed. 
In principle, if the distribution has a 
linear behavior, the slope is insensitive to any smearing due to the resolution. The only 
effect is due to the depletion of the bins at the boundary of the distribution, which 
worsens the $\chi^2$ of the fit. We have verified that the sensitivity to the momentum 
resolution is much smaller for a reduced fit range.

%*****************
\section{Results}
%*****************
About 2 million \semil\ events were selected. The results of the linear fit obtained 
from all run periods are given in Table~\ref{fit1}. The fit is performed separately for \klele\ 
and \klpos\ events to check the reliability of the evaluation of the cluster efficiency. 
The results are consistent only if the respective efficiency corrections for each pion charge 
are applied. Then, combining the two charge results and including the systematic uncertainties 
listed in \Tab{tabsyst} we obtain:

\begin{eqnarray*}
  \lam &=& (28.6\pm 0.5_{\rm{stat.}}\pm 0.4_{\rm{syst.}}) \times \rm{10^{-3}} 
\end{eqnarray*}

The results obtained for the quadratic fit are given in Table~\ref{fit2}. A correlation of 
$\sim-0.95$ between the \lamp\ and \lampp\ parameters is obtained, as expected from the form 
of the parametrization in \Eq{eq:ff2}.
A very slight preference for a small quadratic term is observed as indicated
by the small improvement in the fit probability going from the
linear, P$(\chi^2)=89$\%, to the quadratic fit P$(\chi^2)=92$\%.  
Including the systematic uncertainties listed in \Tab{tabsyst} we obtain:

\begin{eqnarray*}
  \lamp  &=& (25.5\pm 1.5_{\rm{stat.}}\pm 1.0_{\rm{syst.}}) \times \rm{10^{-3}}  \\
  \lampp &=& (1.4\pm 0.7_{\rm{stat.}}\pm 0.4_{\rm{syst.}}) \times \rm{10^{-3}}
\end{eqnarray*}

\begin{figure}[ht]
  \centering
  \begin{tabular}{cc}
    \psfig{figure=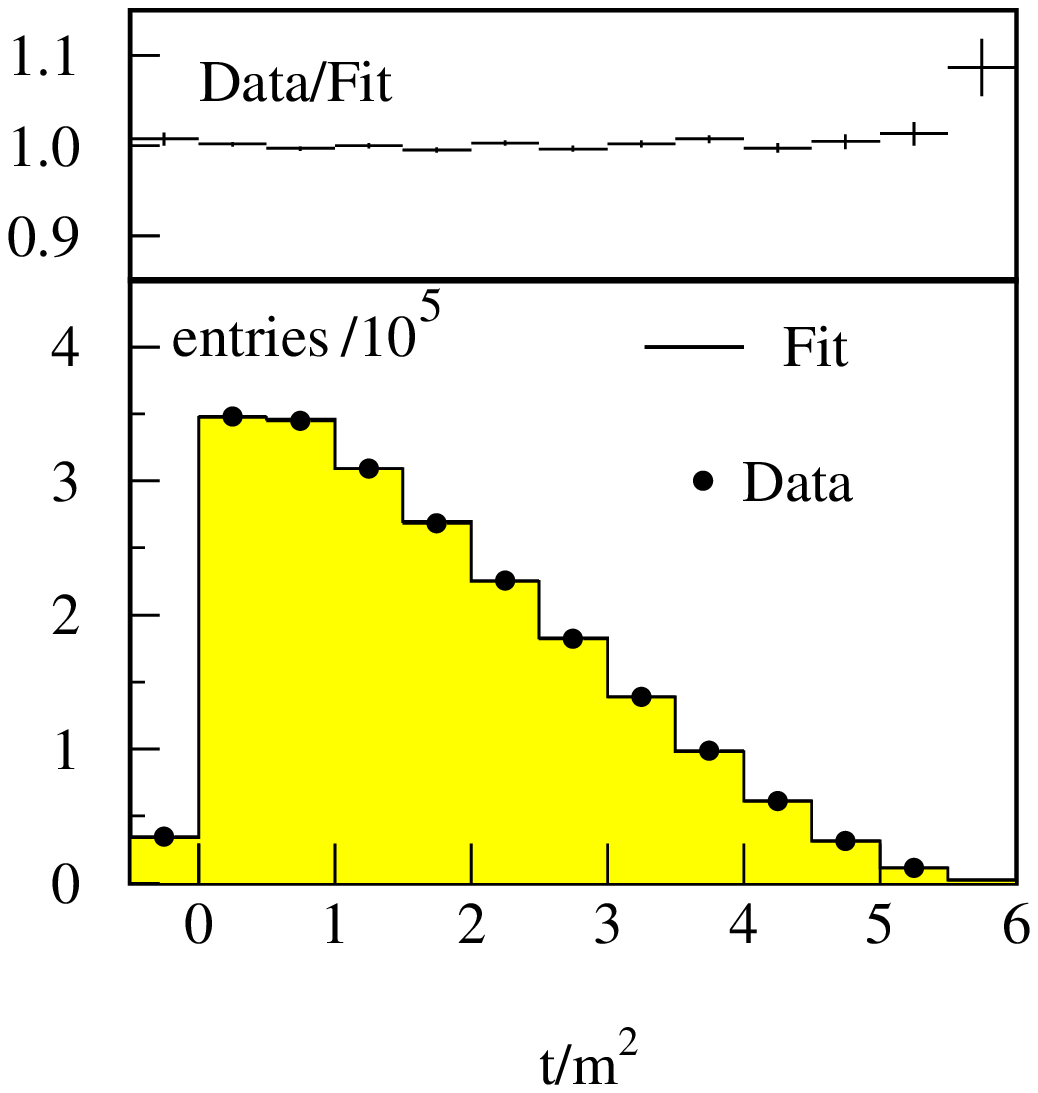,width=6.2cm}
    &
    \psfig{figure=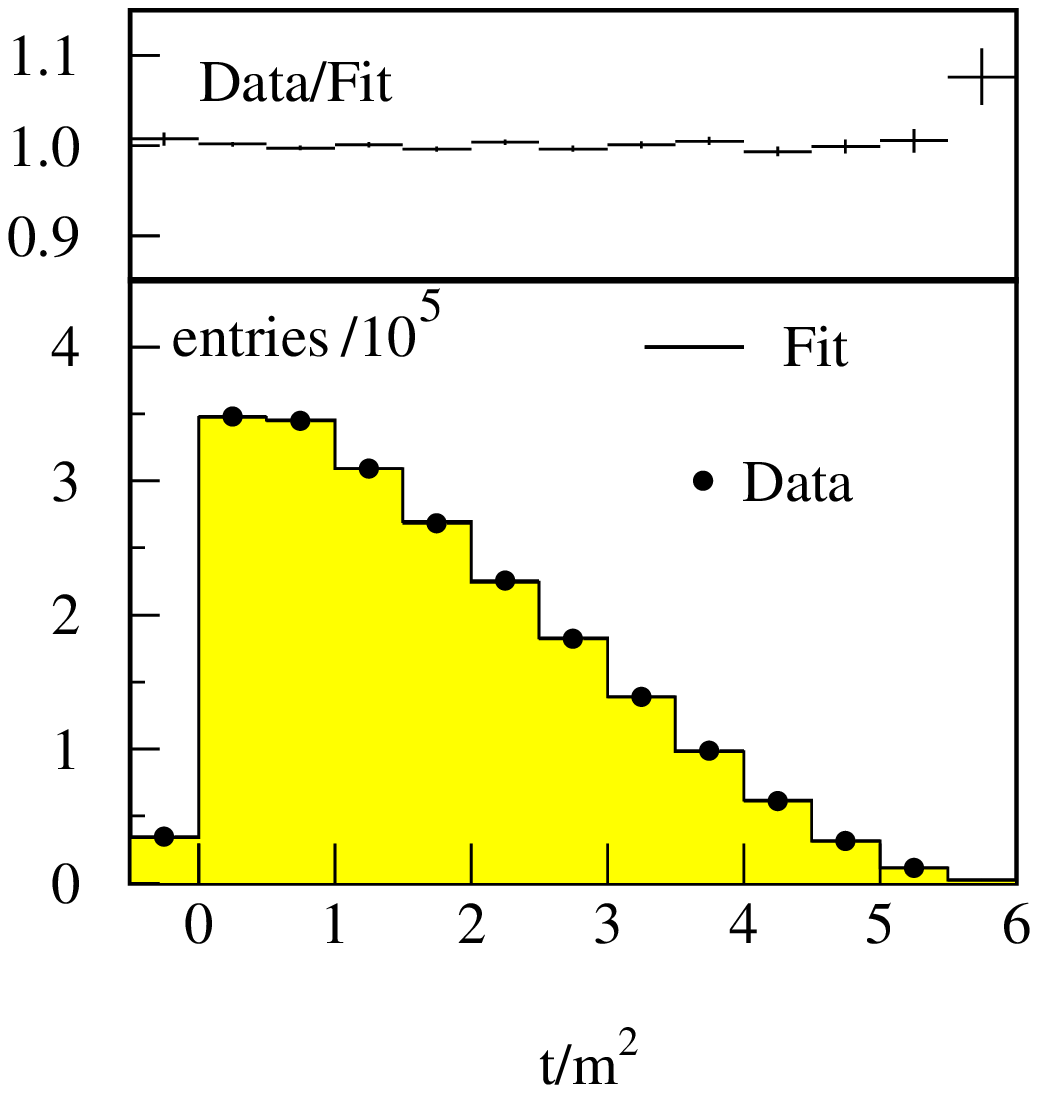,width=6.2cm}\\
    (a) Linear fit & (b) Quadratic fit\\
  \end{tabular}
  \caption{Fit results: data (dots) are superimposed on the fit function (histogram).
    The data/Fit ratio is also shown.}
  \label{fit3}
\end{figure}

Figure~\ref{fit3} shows the $t/m^2$ distribution for the data and the fit
result. The ratio data/fit is also shown.
\begin{table}[hbt]
  \begin{center}
    \begin{tabular}{|c|c|c|}
      \hline
      Linear fit     & $\lam\times10^3$  &  $\chi^2/\rm{ndf}$    \\
      \hline 
      \klpos         & $28.7\pm0.7$       &  $156/181$        \\
      \klele         & $28.5\pm0.6$       &  $174/181$        \\
      Combined       & $28.6\pm0.5$       &  $330/363$        \\
      \hline
    \end{tabular}  
    \caption{Fit results in the hypothesis \lampp=0. Only statistical errors are shown.}
    \label{fit1}
  \end{center}
\end{table}
\begin{table}[hbt]
  \begin{center}
    \begin{tabular}{|c|c|c|c|}
      \hline
      Quadratic fit  & $\lamp\times10^3$ & $\lampp\times10^3$ &  $\chi^2/\rm{ndf}$ \\
      \hline 	     			 					
      \klpos         & $24.6\pm2.1$   & $1.9\pm1.0$     &  $152/180$    \\
      \klele         & $26.4\pm2.1$   & $1.0\pm1.0$     &  $173/180$    \\
      Combined       & $25.5\pm1.5$   & $1.4\pm0.7$     &  $325/362$    \\
      \hline
    \end{tabular}  
    \caption{Fit results with \lamp\ and \lampp\ as free parameters. Only statistical errors 
      are shown.}
    \label{fit2}
  \end{center}
\end{table}
We also fit the data using the one-pole parametrization (see \Eq{eq:pole}).
We obtain   $M_V = (870 \pm 6)\mev$  with $\chi^2/\rm{ndf}=326/364$ and a 
probability of P$(\chi^2)=92.4$\%.
Taking the systematic error into account, we obtain:
\begin{eqnarray*}
  M_V &= &(870 \pm 6_{\rm{stat.}}\pm 7_{\rm{syst.}}) \mev
\end{eqnarray*}
This result indicates that, although the pole is dominated by the $K^*$ vector meson, 
contributions from other $J^P=1^-$ resonant and non-resonant $K\pi$ scattering
amplitudes are not negligible.

%*********************
\section*{Conclusion}
\begin{figure}[ht]
  \centering
  \psfig{figure=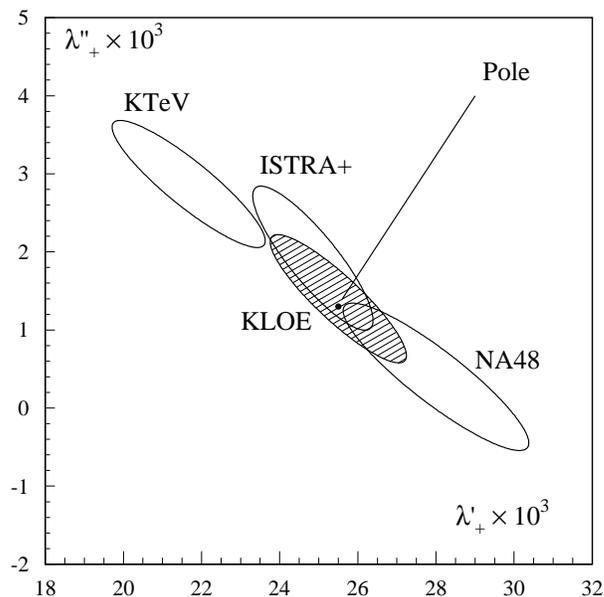,width=8.0cm}
  \caption{Comparison of present results with other recent measurements 
    ~\cite{istra,NA48:ff,KTeV:ff}. The black dot represents the values of \lamp\ and \lampp\ 
    obtained from the Taylor expansion of the pole parametrization. The ISTRA+ result is 
    corrected with the ratio $(m_{\pi^+}/m_{\pi^0})^2$.}
  \label{fig:compa}
\end{figure}

%*********************
We have obtained precise new values of the slopes used to describe the hadronic form factor 
in \semil\ decay. The new KLOE result is consistent with the presence of a small
quadratic term in agreement with the expectation of the one-pole expansion. 
The value of \lam\ obtained with the linear fit is in good agreement with other existing
measurements.
The comparison with other existing measurements is shown in \Fig{fig:compa} in the case of 
the quadratic fit. Our result is in good agreement with ISTRA+~\cite{istra} and NA48 
~\cite{NA48:ff} and in marginal disagreement with KTeV~\cite{KTeV:ff}.
%

%***************************
\section*{Acknowledgements}
%***************************
We thank the DA$\Phi$NE team for their efforts in maintaining low background running 
conditions and their collaboration during all data-taking. We want to thank our technical staff: 
G.F.Fortugno for his dedicated work to ensure an efficient operation of the KLOE Computing Center; 
M.Anelli for his continous support to the gas system and the safety of the detector; 
A.Balla, M.Gatta, G.Corradi and G.Papalino for the maintenance of the electronics;
M.Santoni, G.Paoluzzi and R.Rosellini for the general support to the detector; 
C.Piscitelli for his help during major maintenance periods.
This work was supported in part by DOE grant DE-FG-02-97ER41027; 
by EURODAPHNE, contract FMRX-CT98-0169; 
by the German Federal Ministry of Education and Research (BMBF) contract 06-KA-957; 
by Graduiertenkolleg `H.E. Phys. and Part. Astrophys.' of Deutsche Forschungsgemeinschaft,
Contract No. GK 742; by INTAS, contracts 96-624, 99-37; 
by TARI, contract HPRI-CT-1999-00088.

\bibliographystyle{elsart-num}
\bibliography{paper}

\end{document}